\documentclass[12pt]{article}
\begin{document}
\begin{center}
{\bf Recent Interests on Positron ($e^+$), Positronium (Ps) and Antihydrogen ($\bar H$)} \\ 
\vskip 0.4cm
Hasi Ray 
\footnote{Present address: Positron Laboratory, Department of Physics \& Astronomy, UCR, Riverside, California, USA} 
\vskip 0.3cm
Department of Science, National Institute of Technical Teachers \\
Training and Research (NITTTR) \\
Block FC, Sector 3, Salt Lake City, Kolkata 700106, India \\
Email: hasi\_ray@yahoo.com 
\end{center}
 
\vskip 0.30cm
{\bf Abstract:} A brief survey is made to highlight the recent interests in positron, positronium and antimatter physics. Positron is the first antiparticle observed which was predicted by Dirac. Positronium is itself its antiparticle and bi-positronium molecule is recently observed in laboratory which was predicted by Wheeler in 1946. The simplest antiatom i.e. antihydrogen is observed in the laboratory and the process to achieve the stable confinement of antihydrogen within the trap are in progress to test the standard model. 

\vskip 0.4cm  	

Dirac was awarded the Nobel prize in Physics in 1933 by the Royal Academy of Sciences for developing the basic new ideas of physics, namely his theory of wave mechanics leading upto his relativistic theory of electrons (1928) and holes (1930). Before appearance of Schrodinger's theory, Heisenberg brought out his famous quantum mechanics starting from quite different stand points and viewed his problem from the very beginning with such a broad angle that it took care of systems of electrons, atoms, and molecules. Schrodinger thought that it should be possible to find a wave equation for the motions executed by the electrons which would define these waves in the same way as the wave equation which determined the propagation of light. Although Heisenberg's and Schrodinger's theories had different starting points and were developed by the use of different processes of thought, they produced the same results for problems treated by both theories. Dirac has set up a wave mechanics which starts from the most general conditions. He imposed the condition that the postulate of relativity theory has to be fulfilled. Dirac divided the initial wave equation into two simpler ones, each providing solutions independently. It later appeared that one of the solution systems required the existence of positive electrons having the same mass and charge as the known negative electrons. This initially posed considerable difficulty for Dirac's theory \cite{dirac}, since positively charged particles were known only in the form of the heavy atom nucleus. This difficulty which at first opposed the theory has later become a brilliant confimation of its validity. The existence of the spin of electrons and its qualities are a consequence of this theory. In 1913, Bohr had expressed the idea that Planck's constant should be taken as the determining factor for movements within the atom, as well as, for emission and absorption of light waves. Bohr assumed, after Rutherford, that an atom consists of an inner, heavy, positively charged core, around which negative, light electrons circulate in closed paths, held to the nucleus by Coulomb attraction. Robert Oppenheimer pointed out that an electron and its hole would be able to annihilate each other, releasing energy on the order of the electron's rest energy in the form of energetic photons; if holes were protons, stable atoms would not exist. Hermann Weyl also noted that a hole should act as though it has the same mass as an electron, whereas the proton is about two thousand times heavier. The issue was finally resolved in 1932 when the positron ($e^+$) was discovered by Carl Anderson \cite{anderson}, with all the physical properties predicted for the Dirac hole. \\

The Nobel prize for Physics in 1936 was awarded by V. F. Hess (1/2) for his discovery of cosmic radiation and C. D. Anderson (1/2) for his discovery of the positron. The year 1895 was a turning-point in the history of physics: Rontgen discovered X-rays and this was rapidly followed by Becquerel's discovery of radioactive radiation, and by the discovery of the negative electron by J. J. Thomson (1897) - one of the fundamental elements of atomic structure. Becquerel demonstrated that the radiation emitted by uranium shared certain characteristics with X-rays but, unlike X-rays, could be deflected by a magnetic field and therefore must consist of charged particles. The existence of cosmic radiation became manifest during the search for sources of radioactive radiation. The presence of cosmic radiation offered important problems on the formation and destruction of matter. Carl Anderson, in the course of his comprehensive studies on the nature and qualities of cosmic radiation, succeeded in finding one of the buildingstones of the universe, the positron. Becquerel and Thomson were awarded the Nobel Prizes for physics in 1903 and 1906 respectively for their discoveries. Marie Curie with her husband Pierre Curie, were recognized the Nobel prize in 1901 for their discovery of the radioactive elements radium and polonium. In 1911, Marie Curie (November 7, 1867 to July 4, 1934) was again honored with a Nobel prize, but in chemistry, for successfully isolating pure radium and determining radium's atomic weight. 

 Positron is the first observed antiparticle e.g. antielectron. A Wilson cloud chamber, which is used for detecting particles for ionizing radiation, picture taken by Carl D. Anderson in 1931 showed a particle entering from below and passing through a lead plate; the direction of curvature of the path caused by a magnetic field indicated that the particle was a positively charged one but with the same mass and other characteristics as an electron. The discovery by Anderson, in 1932, of the creation of pairs of electrons and positrons by electromagnetic radiation, and the subsequent interpretation of this observation, in the light of Dirac's already existing relativistic theory of the spinning electron, initiated a fruitful branch of physics which is now often known under the name of "pair theory". The state of disturbance of electron-positron field in the neighbourhood of an atomic nucleus is still imperfectly understood. \\

In 1934, S. Mohorovicic \cite{mohoro} theoretically predicted the existence of the bound system of a positron ($e^+$) and an electron ($e^-$) which is known today as Positronium (Ps), named by Ruark in 1945 \cite{ruark}. There are two types of Ps; one is known as para-Ps and other is ortho-Ps. Para-Ps is a spin singlet state that is in this state of Ps, the spins of positron and electron are antiparallel and it has a life time 125 pico seconds in vacuum. Ortho-Ps is a spin-triplet state, here the spins of positron and electron are parallel and its life time 140 nano seconds in vacuum. Deutsch \cite{deutsch} observed Ps in the laboratory in 1951 in gaseous medium. The distribution of time delays between the emission of a nuclear gamma-ray from the decay of $Na^{22}$ and the appearance of an annihilation quantum had been measured for positrons stopping in a large number of gases and gas mixtures. From the direct observation of the continuous gamma-ray spectrum due to the three-quantum annihilation of triplet positronium in nitrogen confirmed the abundant formation of Ps; since due to electron exchange with the gas molecules having an odd number of electrons, such as nitric oxide, the triplet state of Ps converted very rapidly to the singlet state. \\
 
The $e^+$ being the antiparticle of electron ($e^-$) and Ps being itself its antiatom and the lightest hydrogen-like exotic element, motivated the growing interest of physicists and chemists. Cassidy and Mills \cite{mills1} showed that when intense positron bursts were inplanted into a thin film of porous silica, di-positronium molecule ($Ps_2$) was created on the internal pore surfaces. They found \cite{cassidy} that molecule formation occured much more efficiently than the competing process of spin exchange quenching, observing a reduction in the amount of Ps emitted from an atomically clean Al(1,1,1) surface that depends on the incident positron beam density. If di-positronium is created, then some Ps that might otherwise have been thermally desorbed in the long lived triplet state instead decays at the $Ps_2$ rate of $\sim$ 4 $ns^{-1}$ \cite{schrader}-\cite{bubin}. Since the molecule decays predominantly via two gamma rays while the long lived triplet Ps decays via three photons one could in principle, detect $Ps_2$ using energy selective detectors. The earlier observation of Ps$^-$ \cite{mills2} and the recent observation of Ps$_2$ molecule \cite{mills1} in the laboratory, both the composites were predicted by Wheeler \cite{wheeler} in 1946, have paved the way of further multipositronium work and added a new dimension in antiatom physics \cite{news1}-\cite{news2}.
It is of interest to know the properties of $Ps_2$ \cite{schrader}. This molecule has two electrons and two positrons instead of two protons in hydrogen molecule ($H_2$). All the four constitutents
are of equal masses and it is the lightest molecule. The spin magnetic moment of positron in $Ps_2$ is much more stronger than spin magnetic moment of proton in $H_2$. 
Due to very large spin magnetic moment of positron, the hyperfine structure of Ps becomes comparable to its fine structure. So the spectral behaviours of Ps is expected to be much different from a normal H. Binding energy of di-positronium or $Ps_2$ is E$_b$ = -0.435 eV \cite{varga} while in $H_2$ molecule it is E$_b$ = -4.478 eV. The binding energy of $Ps^-$, E$_b$ = -0.3266 eV \cite{mills2}. Like $H_2$,  $Ps_2$ molecule exists in an overall singlet state \cite{ho}. \\

Wheeler added a note \cite{wheeler} regarding the question of stability of large polyelectrons. According to him, if the stability of the system with two positrons and two electrons i.e. the $Ps_2$ molecule is granted, then the next question regarding the stability comes for such four-particle system i.e. $Ps_4$ \cite{mills3}, when account is taken of the balance between the zero-point kinetic energy of these light masses and the potential energy of van der Waals attraction between them. Soon after the prediction of Wheeler in 1946, Hylleraas and Ore \cite{ore} calculated the binding energy of $Ps_2$. No further work \cite{mezei} on larger polyelectrons appeared in literature. We are trying to calculate the binding energy of a system with four positrons and four electrons. It is expected that $Ps_4$ may have a binding energy, E$_b$ smaller in magnitude than $Ps_2$ because of placing a second $Ps_2$ (see Figure 1) inside a $Ps_2$ (see Figure 2) which may cause a slight reduction in the magnitude of binding energy between the two atoms at the ends of the chain. The symbol $\|$ in figures indicate the electrostatic binding between $e^+$ and $e^-$ in $Ps$. 

\vskip 0.2cm
\begin{center}

$ [\Downarrow e^+]$ \hspace*{0.7cm} $ [\uparrow e^-]$\\

$\| \hspace*{0.6cm} \leftrightarrow  \hspace*{0.6cm} \|$    \\

$[\downarrow e^-]$ \hspace*{0.7cm}    $ [\Uparrow e^+]$ \\
\vskip .2cm
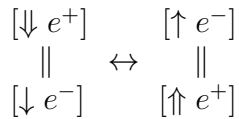
Figure 1. Di-positronium $Ps_2$
\vskip 1cm

$[\uparrow e^-]$ \hspace*{0.7cm} $[\Downarrow e^+]$\hspace*{0.7cm} $[\uparrow e^-]$ \hspace*{0.7cm} $[\Downarrow e^+]$ \\

$\| \hspace*{0.6cm} \leftrightarrow \hspace*{0.6cm} \| \hspace*{0.6cm} \leftrightarrow \hspace*{0.6cm} \| \hspace*{0.6cm} \leftrightarrow \hspace*{0.6cm} \|$   \\

$[\Uparrow e^+]$ \hspace*{0.7cm} $[\downarrow e^-]$ \hspace*{0.7cm} $[\Uparrow e^+]$ \hspace*{0.7cm} $[\downarrow e^-]$\\
\vskip .2cm

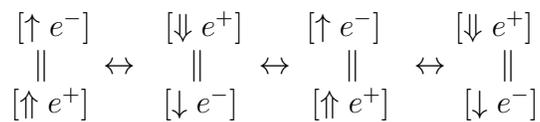
Figure 2. 4-positronium $Ps_4$

\end{center}
\vskip .5cm

\begin{table}[h]
{\bf Table 1. Latest reported binding energies of a few systems:}\\

\begin{tabular}{| c c c | c c c |}  
\hline

 Name&  Symbol    &  Binding Energy &    Name &  Symbol & Binding Energy\\
     &            &  (eV)           &         &         & (eV) \\
\hline

  Positronium & Ps & -6.80 & Hydrogen & H & -13.60  \\

  Di-positronium & $Ps_2$ & -0.43$^a$ & Hydrogen & $H_2$ & -4.48 \\
  molecule      &      &       & molecule &     &        \\

Positronium-ion & $Ps^-$  & -0.33$^b$ & Hydrogen-ion& $H^-$  & -1.05$^c$  \\
                & $Ps^{e^+}$   & ?      &             & $H^{e^+}$    & ?      \\

4-positronium & $Ps_4$ & ? & Positronium    & $PsH$  & -1.06$^d$   \\
molecule      &      &   & Hydride         &     &          \\  
\hline
\end{tabular}
\end{table}
See, $^a$ Ref. \cite{varga}; $^b$ Ref. \cite{mills1}; $^c$ Ref. \cite{drach}; $^d$ Ref. \cite{frolov}. \\

Another remarkable phenomenon of Ps is the formation of Bose Einstein condensate. The Bose-Einstein condensation (BEC) occurs when a macroscopic fraction of an ensemble of particles obeying  Bose statistics collapses into a single state at low temperatures. 
In a non-interacting Bose gas confined by the external harmonic potential $V_{ext}(r) = m (w_x^2 x^2 + w_y^2 y^2 + w_z^2 z^2)/2$, the critical temperature for BEC is given by \cite{albus}
$$ k_B T_c = \hbar \omega_B (\frac{N_B}{\zeta(3)})^{1/3} \simeq 0.94 \hbar \omega_B N_B^{1/3} \eqno(1)  $$
where $\omega_B = (\omega_x\omega_y\omega_z)^{1/3}$ is the geometric mean of the oscillator frequencies, and $m_B$ and $N_B$ are, respectively, the particle mass and the number of bosons in the trap. The above result is obtained using local density approximation (LDA), where the temperature of the gas is assumed to be much larger than the spacing between single particle levels: $k_BT \gg \hbar \omega_{x,y,z}$. In this case the density of thermal atoms can be written as 
$$n_B({\bf r}) = (\lambda_T^B)^{-3} \sum_{n=1}^{\infty}\frac{(e^{-[V_{ext}^B({\bf r}) - \mu_B]/k_BT})^n}{n^{3/2}} \eqno(2) $$
where $\lambda_T^B = h/(2m_Bk_BT)^{1/2}$ is the boson thermal wavelength. At $T = T_C$ the boson chemical potential takes the critical value $\mu_B = \mu_C = 0$, corresponding to the bottom of the external potential, and the density $n_B(0)$ in the centre of the trap satisfies the critical condition $n_B(0)(\lambda_T^B)^3 = \zeta(3/2) \simeq 2.61$ holding for a homogeneous system.
Here 'h' is Planck's constant, `$k_B$` is Boltzmann constant and $\hbar = h / 2\pi$ . 
As the temperature is lowered below T$_c$ the number of particles in the zero momentum state $ < n_0 >$ develops a macroscopic value \cite{mills4}:
$$ < n_0 > /n = 1 - (T/T_c)^{3/2}    \eqno(3) $$
 is comprised of an $e^+ - e^-$ bound in a hydrogenic orbit. Its mass, 2$m_e$, is extremely light compared to H, an important ingredient \cite{adhikari} for achieving reasonable Bose condensation temperatures. 
As a purely leptonic, macroscopic quantum matter-antimatter system this would be of interest in its own right, it would also represent a milestone on the path to produce an annihilation gamma-ray laser.\\

In addition, the first confirmed production of cold antihydrogen ($\bar H$) atoms in a confinement trap \cite{charlton1} in 2002 and the initiative to achieve the stable confinement of neutral atoms within the trap has created a considerable excitement to both the physicists and chemists.  $\bar H$ is an ideal system for testing the standard model \cite{book} - \cite{stan-web} prediction of the symmetry between matter and antimatter. According to this model, systems made up of antimatter should behave identically to those composed of matter. 
Just as a hydrogen atom ($H$) consists of an electron orbiting a proton, an antihydrogen atom ($\bar H$) consists of a positron orbiting an antiproton. The guessing is that the sprectrum of $\bar H$ looks exactly like that of $H$. After all, the emission spectrum of H is due to an excited electron jumping from the excited energy level down to a lower level(s): presumably the positron in $\bar H$ has the same separation of energy levels. Any difference between the emission spectrum in $H$ and $\bar H$ would be a new indication. The use of laser spectroscopy to measure and compare the electronic structures of $\bar H$ to that of normal $H$ (e.g. antiatom and atom) is a new and an interesting area \cite{web1}. \\

The spin polarized atomic hydrogen H$\downarrow$ (i.e. both the proton and electron have the down spins) or H$\uparrow$ (i.e. both the proton and electron have the up spins) is expected to form ${\it no}$ molecule and it will remain a gas (in the atomic state) down to zero temperature. At densities $n \cong 10^{16} cm^{-3}$ the system is weakly interacting and will Bose condense at temperatures of roughly 10$^{-2}$ K. Although a gas of H$\downarrow$ or H$\uparrow$is a good approximation to an ideal Bose system, workers have been unable to achieve high enough densities or low enough temperatures to observe its Bose condensation \cite{mills4}. The exciton gas produced by pumping an insulator like Cu$_2$O with a short laser pulse is also a promising candidate. This system is in many ways very analogous to the positronium system we discuss here.  
A collection of the spin polarized ortho Ps (Ps$\downarrow$ i.e. both the positron and electron have the down spins or Ps$\uparrow$ i.e. both the positron and electron have the up spins) seems to be viable in the laboratory to achieve high temperature BEC. The critical temperature T$_c$ for BEC of ideal bosons of mass $m$ and density $n$ is given by $ T_c = (h^2/2mk_B)(n/2.61)^{2/3}$. Hence the small mass of Ps at a very low density, $n = 10^{12} cm^{-3}$ should facilitate BEC by leading to a large T$_c \sim 10^{-2}$ K.\\

 According to Dirac's theory, antimatter particles should have the same mass, but opposite charge as their matter equivalents e.g. the simplest antimatter of electron is positron. Antimatter is naturally formed during the radioactive decay of some elements. However, such naturally occurring antimatter is too little to be able to produce significant collection of the system. They are quasi stationary system i.e. their life time is very short ($\sim$ pico seconds) - this period of time proves inadequate for collection and experimentation. This had led to the need for further research and study on how to produce large amounts of antimatter under controlled conditions.\\

Antimatter can be used to sensitively test the theoretical underpinnings of the standard model.  Essential to the quantum field theory governing interactions of fundamental particles is the so-called CPT theorem, which involves discrete symmetries.  The CPT theorem requires that the laws of physics be invariant under the following operation: all particles are replaced by their antiparticle counterparts (charge conjugation), all spatial coordinates are reflected about the origin (parity), and the flow of time is reversed (time reversal).  The CPT theorem has important implications for antimatter, including the mass equivalence of particle and antiparticle. \\

Another reason why $\bar H$ is worth studying is its potential to test the weak equivalance principle (WEP) of Einstein's general relativity, which requires the gravitational acceleration of a falling body be independent of its composition. This has been tested rigorously for different objects of matter, but tests of antimatter and direct comparison of a matter object and its antimatter equivalent, such as protons and antiprotons, have proved very difficult, mainly due to the difficulty of shielding for even very small electromagnetic fields. This is necessary since the elctromagnetic force is much stronger than gravity. $\bar H$, on the other hand, is thought to be stable and neutral and tests using this should thus be enabled at much higher accuracy. Slow neutral $\bar H$ suitable for a free fall measurement, is currently being proposed by Walz and H$\ddot{a}$nsch; the laboratory of CEA/Sacley, France \cite{walz} is presently engaged in producing the slow neutral $\bar H$ needed for this experiment. They have proposed the use of $\bar H^+$ ion in order to collect ultra cold $\bar H$ \cite{perez}. For this a dense Ps target is necessary to follow up: $\bar p$ + Ps $\rightarrow$ $\bar H$ + $e^-$ which is followed by $\bar H$ + Ps $\rightarrow$ $\bar H^+$ + $e^-$. This $\bar H^+$ ion could be cooled to $\mu$K temperatures (i.e. m/s velocities). The excess positron can be laser detached in order to recover neutral $\bar H$. \\

The precision tests of CPT invariance using antimatter include the electron/positron mass ratio and the proton/antiproton mass ratio.  An ideal system for more precise studies of the CPT theorem is the antihydrogen atom.  The CPT theorem requires that hydrogen and antihydrogen have the same spectrum.  Since hydrogen is one of the best understood and most precisely studied systems in all of physics, it is natural to try to compare the spectra of hydrogen and antihydrogen.\\

Whenever antimatter collides with its equivalent matter, they will annihilate each other. This collision and annihilation will release large amounts of energy because in the process, the mass of both particle and antiparticle will be converted into pure energy - usually in the form of high-energy photons (known as gamma rays). The energy be released from such a collision (according to Einstein's equation, `$E=mc^2$`) could be used to generate electricity using advanced technology and equipment.\\

Intensive studies are currently being undertaken by numerous institutions regarding the behavior and application of the antimatter.
CERN's unique new antimatter factory, the Antiproton Decelerator (AD) has begun delivering antiprotons to experiments. These experiments will study antimatter in depth to determine if there is a difference between it and ordinary matter. Any difference between antimatter and matter would be extremely interesting since it is not yet understood why the universe is made mostly of matter. Physicists believe that the Big Bang created equal amounts of antimatter and matter \cite{book1}, which would then have annihilated, leaving nothing. The great mystery is why there was enough matter left over to from the universe. Two experiments, ATHENA \cite{athena} and ATRAP \cite{atrap}, aim to add positrons - anti-electrons - to the caged antiprotons to make atoms of antihydrogen. A third, ASACUSA \cite{asacusa}, traps the antiprotons in a cage conveniently provided by nature – the helium atom. The goal of all three is a detailed comparison of matter and antimatter leading to an understanding of why nature has a preference for matter over antimatter.\\

 As technology advances through the years, better and cheaper ways of producing significant amounts of antimatter are expected to be developed and antimatter may become a good source of renewable and sustainable energy. This is not yet possible today, but in the future, this might be a significant possibility.\\

Author is thankful to Saha Institute of Nuclear Physics (SINP), Kolkata for providing the facilities to prepare the manuscript and grateful to Prof. Bichitra Ganguly (SINP) for her active helpful suggestions to improve it. Author is thankful to SERC, DST, Govt. of India for the project funding Ref. No. SR/WOSA/PS-13/2009 and Prof. Anuradha De (NITTTR) for her kind mentorship.

\end{document}